\documentclass[twocolumn,prd,noshowpacs,nofootinbib,amsmath,amssymb,superscriptaddress]{revtex4}
\usepackage{graphicx,epsfig,psfrag,bm,amssymb}
\usepackage{dcolumn}
\usepackage{bm}
\usepackage{mciteplus}
\usepackage{color}
\usepackage{mathrsfs,amsfonts,hepunits, color}
\usepackage{mciteplus}
\usepackage{tikz}
\usepackage{slashed}
\usetikzlibrary{arrows,shapes}
\usetikzlibrary{trees}
\usetikzlibrary{matrix,arrows} 				% For commutative diagram
\usetikzlibrary{positioning}				% For "above of=" commands
\usetikzlibrary{calc,through}				% For coordinates
\usetikzlibrary{decorations.pathreplacing}  % For curly braces
\usepackage{pgffor}							% For repeating patterns

\usetikzlibrary{decorations.pathmorphing}	% For Feynman Diagrams
\usetikzlibrary{decorations.markings}
\usetikzlibrary{snakes}

\tikzset{
    vector/.style={decorate, decoration={snake}, draw},
	provector/.style={decorate, decoration={snake,amplitude=2.5pt}, draw},
	antivector/.style={decorate, decoration={snake,amplitude=-2.5pt}, draw},
    fermion/.style={draw, postaction={decorate},
        decoration={markings,mark=at position .55 with {\arrow[draw]{>}}}},
    fermionbar/.style={draw, postaction={decorate},
        decoration={markings,mark=at position .55 with {\arrow[draw=black]{<}}}},
    fermionnoarrow/.style={draw},
    gluon/.style={decorate, draw,decoration={coil,amplitude=4pt, segment length=6pt}, line width=1},
    scalar/.style={dashed,draw, postaction={decorate},
        decoration={markings,mark=at position .55 with {\arrow[draw]{>}}}},
    scalarbar/.style={dashed,draw, postaction={decorate},
        decoration={markings,mark=at position .55 with {\arrow[draw]{<}}}},
    scalarnoarrow/.style={dash pattern = on 6 pt off 3 pt,draw},
    electron/.style={draw, postaction={decorate},
        decoration={markings,mark=at position .55 with {\arrow[draw]{>}}}},
	bigvector/.style={decorate, decoration={snake,amplitude=4pt}, draw},
	vectorscalar/.style={loosely dotted,draw, postaction={decorate}},
}
\RequirePackage{xspace}
\usepackage{relsize}
\usepackage{slashed}

\newcommand{\be}{\begin{eqnarray}}
\newcommand{\ee}{\end{eqnarray}}
\def\lsim{\mathrel{\rlap{\lower4pt\hbox{\hskip 0.5 pt$\sim$}}
    \raise1pt\hbox{$<$}}}                % less than or approx. symbol
\def\gsim{\mathrel{\rlap{\lower4pt\hbox{\hskip1pt$\sim$}}
    \raise1pt\hbox{$>$}}} 
    
\def\lsim{\mathrel{\rlap{\lower4pt\hbox{\hskip1pt$\sim$}}
    \raise1pt\hbox{$<$}}}
\def\gsim{\mathrel{\rlap{\lower4pt\hbox{\hskip1pt$\sim$}}
    \raise1pt\hbox{$>$}}}

\newcommand{\gev}{{\rm GeV}}

\begin{document}

\title{Big Bang  Darkleosynthesis}
\author{Gordan Krnjaic}
 \affiliation{Perimeter Institute for Theoretical Physics, Waterloo, Ontario, Canada    }
\author{Kris Sigurdson}
 \affiliation{Department of Physics and Astronomy, University of British 
Columbia, Vancouver, BC, Canada}

\begin{abstract}
In a popular class of models, dark matter comprises an asymmetric population of 
composite particles with short range interactions arising from a confined nonabelian gauge group. 
We show that coupling this sector to a well-motivated light mediator particle 
yields efficient {\it darkleosynthesis}, a dark-sector version of big-bang nucleosynthesis (BBN), in generic regions of parameter space.
   Dark matter self-interaction bounds typically require the confinement scale to be above 
  $\Lambda_{QCD}$, which generically yields large ($\gg$ MeV/dark-nucleon) binding energies. 
These bounds further suggest the
 mediator is relatively weakly coupled, so repulsive forces between 
 dark-sector nuclei are much weaker than coulomb repulsion between standard-model nuclei, which
results in an exponential barrier-tunneling enhancement over standard BBN.   
Thus, {\it darklei} are easier to make and harder to break than visible 
   species with comparable mass numbers.
 This process can efficiently yield  a dominant population of states 
  with masses significantly greater than the confinement scale and, in contrast to dark matter 
  that is a fundamental particle, may allow the dominant form of dark matter to have high spin ($S \gg3/2$), whose discovery would
  be smoking gun evidence for dark nuclei. 
\end{abstract}

\maketitle

%%%%%%%%%%%%%%%%%%%%%%%%%%%%%%%%%
%%%%%%%%%%%%%%%%%%%%%%%%%%%%%%%%%
%
%					Introduction
%
%%%%%%%%%%%%%%%%%%%%%%%%%%%%%%%%%
%%%%%%%%%%%%%%%%%%%%%%%%%%%%%%%%%

\section{Introduction}
There is abundant evidence for the existence of dark matter (DM) but its particle nature is still unknown \cite{Beringer:1900zz}. 
A popular, well motivated class of models \cite{Alves:2009nf,Alves:2010dd, Buckley:2012ky, Lee:2013bua, Kribs:2009fy,Khlopov:2010pq}
features a composite dark sector with asymptotically free confinement and a matter asymmetry   
  in analogy with standard model (SM) quantum chromodynamics (QCD).
 At temperatures below the confinement scale $\Lambda_D$, this sector 
 comprises hadron-like particles with short-range 
 self-interactions and requires no ad-hoc discrete or global symmetries to protect its
 cosmological abundance from decays. 
 
 In this paper we consider the implications of big bang {\it darkleosynthesis} (BBD) -- the synthesis of {\it darklei} (dark-sector nuclei)
 from {\it darkleons} (dark-sector nucleons) in the early universe --- in an asymmetric nonabelian sector coupled to 
 a lighter ``mediator" ($m_{\rm med.} \ll \Lambda_D$) particle. A mediator is well motivated
 in asymmetric DM as it facilitates annihilation in the early universe to 
avoid a higher than observed dark-matter abundance  \cite{Feng:2008ya,Izaguirre:2013uxa} and allows for DM self interactions, which can resolve  
 puzzles in simulations of large scale structure formation \cite{Kaplinghat:2013kqa}, and may explain anomalies
  in direct and indirect detection experiments (see \cite{Feng:2010zp, Gresham:2013mua} and references therein).  

%%%%%%%%%%%%%%%%%%%%%%%%%%%%%%%%%
%					FIG Avg Mass Contour
%%%%%%%%%%%%%%%%%%%%%%%%%%%%%%%%%
\begin{figure}[h!] 
 \vspace{0.cm}
 \begin{center}
 \hspace{-0.5cm} \includegraphics[width=7.1cm]{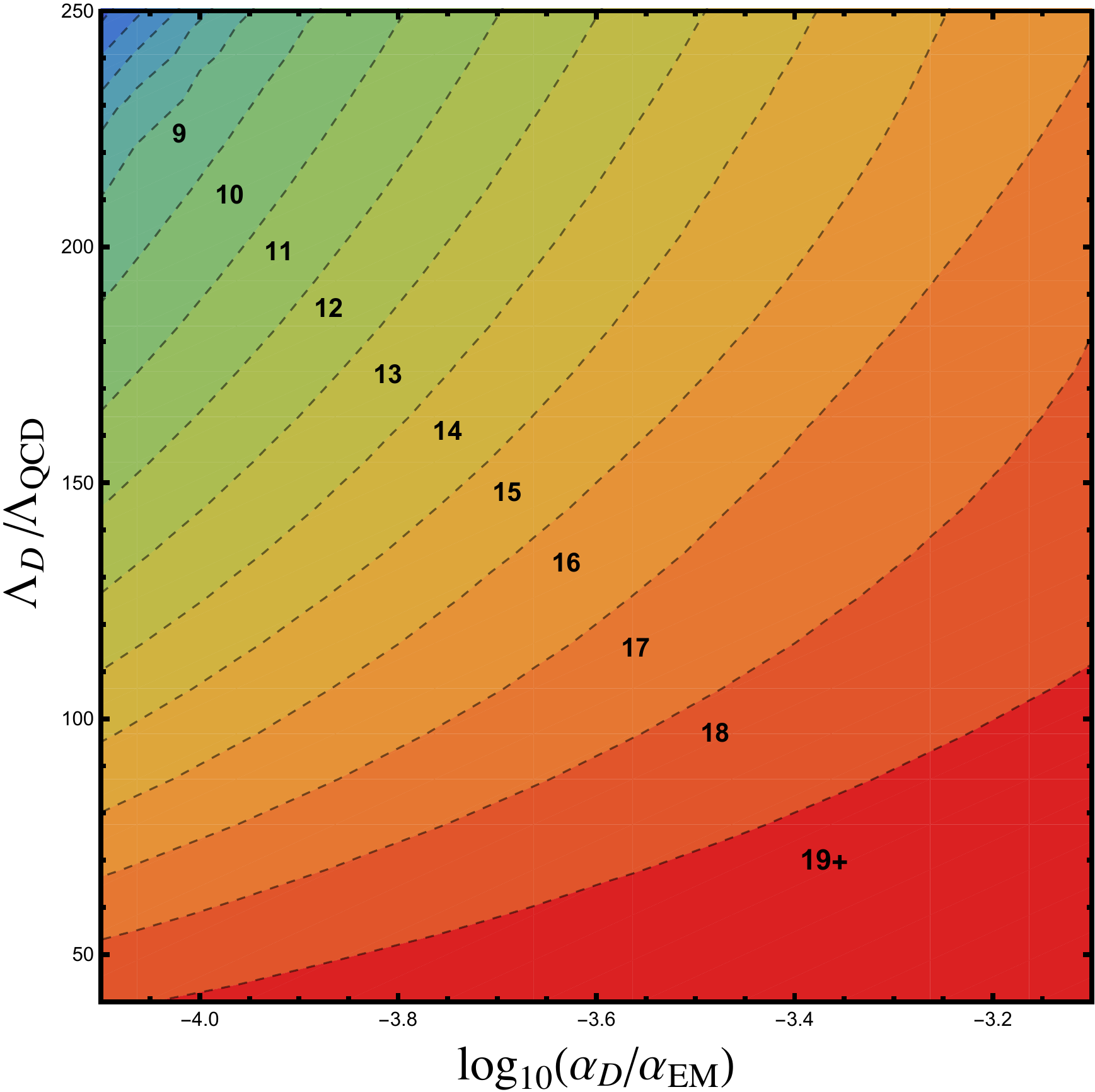} 
  \caption{ Expected dark-nuclear mass-number $ \langle A \rangle \equiv \sum_A  A^2 Y_A $ for a
 from an initial population of identical $\chi$ darkleons with masses $m_\chi  \equiv m_n (\Lambda_D/\Lambda_{QCD})$ and a
   $U(1)_D$ gauge boson $V$ whose mass is negligible during BBD. The simulation solves 
   the Boltzmann equations in Eq.~\ref{eq:boltz} for all species up to $A_{max} = 20$
   using the binding model in Eq~\ref{eq:binding} with parameter values described in the text.
 Note that for this binding model,
 the most tightly bound species has $A^* \gg A_{max}$; as a proof of principle, our simulation 
truncates the species for numerical tractability even though much higher composites would likely form in
this setup. This truncation combined with small $\alpha_D \ll \alpha_{EM}$ approximates the 
behavior of a system with larger $\alpha_D$ for which $ \langle A \rangle \sim A^*\sim 20.$   
 Constraints from DM self interactions $\sigma/m_\chi \lsim 0.1 - 1$ cm$^2$/g \cite{Vogelsberger:2012ku, Randall:2007ph} are trivially 
 satisfied for this parameter space.
 For comparison with visible-sector BBN, we have also numerically verified that making
  species $A = 5, 8$ unstable {\it by fiat} preserves the qualitative character of these results,
  demonstrating that high-occupancy darklei can efficiently form 
 through $\Delta A > 1$ reactions.
 }
  \end{center}\label{fig:contours}
\vspace{-0.4cm}
\end{figure}
  
%%%%%%%%%%%%%%%%%%%%%%%%%%%%%%%%%
%				      FIG: Yield Plot 
%%%%%%%%%%%%%%%%%%%%%%%%%%%%%%%%%
\begin{figure*}[t]
 \begin{center} 
 \vspace{-0.2cm}
 \includegraphics[width=8cm]{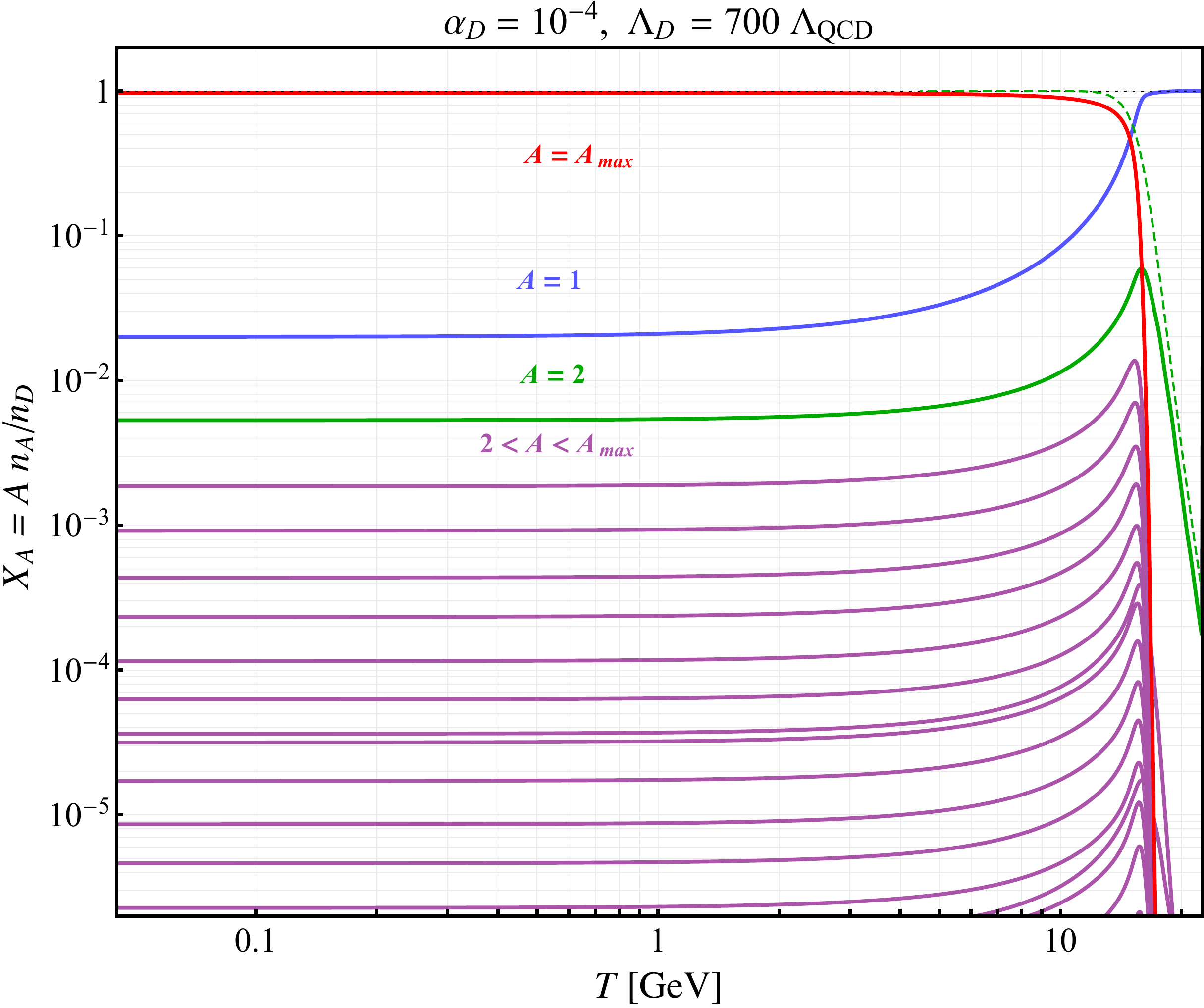}  ~~
 \includegraphics[width=8cm]{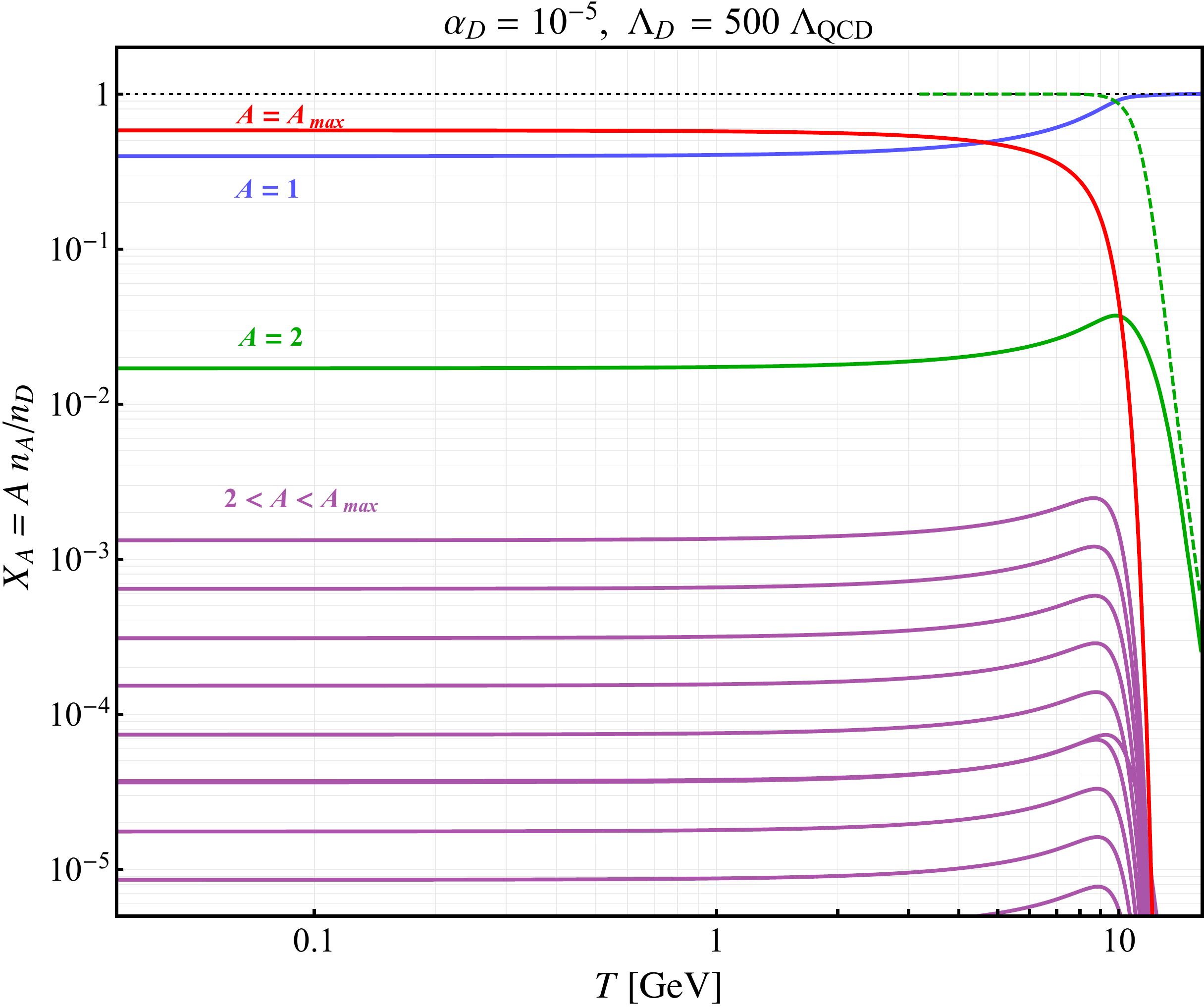}  
  \caption{
 Example mass fractions $X_A \equiv A n_A / n_{D}$  for various inputs with darkleon mass
  $m_\chi = m_n (\Lambda_D/\Lambda_{QCD})$ computed by solving the Boltzmann system in Eq.~\ref{eq:boltz}. The blue curve (color online) in each plot is the free darkleon fraction $(A=1)$, 
 the red curve $(A = A_{max})$ is the maximum occupancy number included in the simulation, and the green curves are is the dark ``deuterium"  ($A = 2$) 
 mass fraction computed in both the Saha approximation (dashed) and the full Boltzmann solution (solid).
  The purple curves are number fractions for all other species. We simulate all species $A = 1-20$  for each data point, but our 
  results are qualitatively similar when species $A = 5, 8$ are removed (as in conventional, visible sector BBN). 
  All plots assume the binding model parameters described in the text and, to be conservative, we 
 also assume all $A>2$ species start with zero abundance and solve the system out to final temperature $T ={\cal B}(2)/1000$.
 The initial condition for $A = 2$ is set by the Saha solution at initial temperature $T =  {\cal B}(2)/20$.   
   }
  \end{center}\label{fig:yields}
\vspace{-0.4cm}
\end{figure*}

In the limit where the mediator is sufficiently
weakly coupled, the initial conditions 
in the dark sector are analogous to those considered in the ``alphabetical article"
by Alpher and Gamow \cite{Alpher:1948ve}, who sought to build up 
all the observed chemical elements from only an initial population of SM neutrons during big bang nucleosynthesis (BBN).
Although this proposal ultimately failed as an efficient and complete model of nucleosynthesis, we show here that this need
not be the case when considering the build up of darklei from darkleons.
 In the dark
sector, such a setup can be realized more generically and need not
encounter the (perhaps accidental) coincidences (e.g., $m_n-m_p \sim T_{BBN}$) that prevent visible BBN from building up 
species with large mass numbers.  Though other work has considered the cosmology of dark-sector bound states via dark
``recombination" \cite{MarchRussell:2008tu,MarchRussell:2008yu,Kaplan:2009de,Kaplan:2011yj,CyrRacine:2012fz,Cline:2013zca,Cline:2013pca,Cline:2012is, Behbahani:2010xa},
and in the context of mirror matter \cite{Foot:2004pa, Foot:2014mia, Ciarcelluti:2014vta}, 
to our knowledge this is the first demonstration that dark-sector nucleosynthesis is a generic possibility for confined dark matter 
scenarios. 
 
 We assume only that the dark sector is populated with self-interacting, non-annihilating
 darkleons that also couple to a light mediator, which enables di-darkleon formation; 
 in the absence of this coupling, there is no available energy loss mechanism for di-darkleon formation.
Although in principle the coupling to the mediator state can induce both attractive and repulsive interactions between darklei, 
to be conservative and demonstrate viable phenomenology despite coulomb repulsion, we assume here that all dark-nuclear
 formation rates feature repulsive barriers.  

 The outline of this paper is as follows. Section \ref{sec:basic-ingredients} 
 outlines the basic ingredients of our scenario, Section~ \ref{sec:uv} outlines
a concrete UV complete realization, and Section~\ref{sec:conclusion} offers
 some concluding remarks and speculations.

%%%%%%%%%%%%%%%%%%%%%%%%%%%%%%%%%
%%%%%%%%%%%%%%%%%%%%%%%%%%%%%%%%%
%
%					Basic Ingredients 
%
%%%%%%%%%%%%%%%%%%%%%%%%%%%%%%%%%
%%%%%%%%%%%%%%%%%%%%%%%%%%%%%%%%%

\section{ Basic Ingredients}\label{sec:basic-ingredients}
{\it\bf Nonabelian Sector} Our starting point is to consider a matter asymmetric dark sector with a single species of fermionic dark-``quarks" 
 charged under an $SU(N)$ gauge group. We assume this group becomes confining at some scale $\Lambda_D$
 at which the quarks form darkleons $\chi$. In the simplest scenario, the darkleon mass comes predominantly
 from strong dynamics, so the constituent quark masses can be neglected. However, we assume them to be nonzero,
 so if an approximate chiral symmetry is broken by confinement, 
 the dark ``pions" will be massive and decay to the visible sector through the mediator described below. 

%%%%%%%%%%%%%%%%%%%%%%%%%%%%%%%%%
%				  	Light Mediator 
%%%%%%%%%%%%%%%%%%%%%%%%%%%%%%%%%
{\it \bf Light Mediator:}
To demonstrate nucleosynthesis in the dark sector, we couple our darklei to a lighter 
particle $V$  that enables di-nucleon formation $\chi + \chi \to  {^{2}\chi} + V$, where $^A\chi$ denotes
a darkleus with mass number $A$; in the absence of $V$ emission this process is kinematically forbidden. 
Furthermore, in order for any dark matter scenario to have observable consequences, there needs to be an operator
that connects dark and visible sectors. 

Both problems can be solved with a light mediator particle uncharged under the confining gauge group. 
One well motivated example identifies the mediator $\phi$ with a kinetically-mixed $U(1)_D$ gauge boson \cite{Holdom:1985ag} $V$ whose 
lagrangian is
\be
{\cal L} =   \frac{\epsilon}{2} F^\prime_{\mu\nu}F^{\mu\nu} + \frac{m^2_{V}}{2} V_\mu V^{\mu} + 
 \bar \chi (i \gamma^\mu D_\mu + m_\chi) \chi  ~~,
%  \\ D_\mu = \partial_\mu  +i g_D Z_\chi V_\mu ~~~~,~~~~~~~~~~~~~~~~~
\ee
where $F^\prime_{\mu\nu} \equiv \partial_{[\mu,} V_{\nu]}$ is its field strength, $m_V$ is its mass, $\alpha_D$ is the dark fine-structure constant, 
 and  $\chi$ is a dark-nucleon with $U(1)_D$ charge $Z_\chi$ and mass $m_\chi$. Independently of the connection
  to BBD this mediator can resolve the persistent 
 $(g-2)_\mu$ anomaly \cite{Pospelov:2008zw}. Phenomenologically, 
 $V$ must decay before visible BBN, which can easily accommodated 
in our regime of interest $\Lambda_D \gg \Lambda_{QCD}, m_{V}$  \cite{Essig:2013lka}, where $\Lambda_{QCD} = 200$ MeV. 

In a matter asymmetric sector, $U(1)_D$ charge neutrality
requires at least one additional species of with opposite charge, which yield a variety of net nuclear charges
after darkleosynthesis. We will return to this possibility in Sec.~\ref{sec:uv}, but note that having identical,
repulsive charges under the mediator is a conservative choice that yields the maximum repulsion between
fusing species to suppress formation rates.

%%%%%%%%%%%%%%%%%%%%%%%%%%%%%%%%%
%					Binding Model
%%%%%%%%%%%%%%%%%%%%%%%%%%%%%%%%%

{\it \bf Binding Model:}
 In the visible sector, the liquid drop model \cite{Baym:1971ax, Baym:1977} gives the approximate binding energy for a species 
with mass number $A$ 
\be\label{eq:binding}
{\cal B}(A) &=& a_V A - a_S A^{2/3} -   a_C Z^2 A^{-1/3}  \nonumber  - \delta(A)~~,  
%\\ &-&  a_A A^{-1}(A-2Z)^2 
\ee
where $a_V, a_S$, and $a_C,$ are respectively the volume, surface, and coulomb terms,
while $\delta(A) =\pm a_P A^{-1/2}$ is the pairing term with $+(-)$ for $A$ odd (even).
Since we will only consider a single-species of dark-nucleon, we neglect isospin by setting $A=Z$ in
the familiar parametrization. 

Some  intuition into the physical relationship between these coefficients in this binding model can be gained by calculating the Yukawa self-energy ${\cal B}_{SE}$ of a uniform-density sphere in an effective theory of nuclear reactions mediated by pion-like scalars of mass $m_{\Pi}$.  For finite $m_{\Pi} \sim O(\Lambda_D)$ it is straightforward to show that ${\cal B}_{SE} \simeq \kappa_V (\Lambda_D^3/m_{\Pi}^2)A-\kappa_S (\Lambda_D^4/m_{\Pi}^3)A^{2/3}$, and thus the relative importance of the surface term compared to the volume term depends on the range of nuclear forces (parametrized as $\Lambda_D/m_{\Pi}$).

We adopt this simple model (along with the notation) to calculate dark nuclear binding energies for species with mass number $A$
and interpret $Z$ to be the number of constituents with unit charge under $U(1)_D$.
Note that for large $A$, where $\delta \ll a_S,a_C$ the most tightly bound 
species has mass number $A^* \simeq a_S/2a_C$, which maximizes
 $ A^{-1}{\cal B}(A)$, the binding energy per darkleon.  We note for a fixed volume term smaller $A^*$ occur for a shorter nuclear range (smaller $a_S$) or larger $\alpha_D$ (larger $a_C$). For our numerical studies, we take the inputs $a_V, a_S,$ and $a_P$ to be of order the 
confinement scale $\Lambda_D$, but take the  $U(1)_D$ Coulomb term to be parametrically 
smaller to reflect the absence of long range self interactions at late times.  
 
%%%%%%%%%%%%%%%%%%%%%%%%%%%%%%%%%
%					Cross Sections
%%%%%%%%%%%%%%%%%%%%%%%%%%%%%%%%%

{\it  \bf Formation \& Destruction Rates:}
The generic ``strong" reaction involving  species $A$ and $B$ with net darkleon transfer $C$  
is $^A\chi +\,  ^B\chi \to\, ^{(A+C)}\chi +\, ^{ (B-C)}\chi$. 
We adopt the prescription in \cite{Wagoner:1966pv} (and references therein) 
to parametrize the strong cross-section for this process as
\be
\sigma(E; A, B) = \frac{(A^{1/3} + B^{1/3})^2}{\Lambda^2_D} ~e^{-F(A, B)/E^{1/2}} ~~,
\ee
where $E$ is the {\it kinetic} energy and $F(A,B) \equiv  \alpha_D A B  (2 \mu)^{1/2}$ is the coulomb-barrier 
tunneling coefficient for
 repulsive $U(1)_D$ interactions between initial-state particles, and $\mu$ is their reduced mass. 
In the $\alpha_D \to 0$ limit, this expression recovers the geometric scattering limit\footnote{ Although the cross section increases with the $A$ and $B$, this ansatz
 never violates self-scattering unitarity bounds because the  cross section
in Eq.~2 is of the form $\sigma  \sim R^2$, where $R\sim A^{1/3} / \Lambda_D$ is the 
radius of an incident nucleus, so the bound 
\cite{Griest:1989wd} on geometric cross sections is $\sigma \lsim 16 \pi R^2$ weakens for larger objects  if other
couplings are perturbative.}.
Thermal averaging with the Maxwell-Boltzmann distribution yields
%\footnote{The lower integration limit is 
%set by the range of the strong force, namely $m_\Pi$, which is parametrically related to $\Lambda_D$. However, for geometric 
%cross sections, integrating from zero induces only a negligible correction and can be neglected.}
\be 
\label{eq:thermal-integral}
\hspace{-0.2cm}
\langle \sigma v \rangle_{A,B} =  \frac{2 (A^{1/3} + B^{1/3})^2 }{\sqrt{\pi} \Lambda^2_D \,T^{3/2}} \int_0^\infty  \! dE E^{1/2}  v(E)
e^{-U(E,T; A,B)}\!,~
\ee
where the angle-averaged relative velocity between fusing species is $v(E)  = \sqrt{  v^2_{A} +v^2_{B}   }$, 
$v_{i} = \sqrt{1 - m_i^2/(m_i + E)^2}$ is the center-of-momentum velocity for species $i$ and 
\be
U(E,T; A,B) = E/T + F(A,B)/E^{1/2} ~~,
\ee
includes the usual Boltzmann factor and coulomb barrier. Although we include the latter
for completeness (and for comparison with standard BBN), we always
work in the regime $\alpha_D \ll 1, F(A,B)/\sqrt{T} \ll 1$, so this correction is negligible and our interactions
are  thermally-averaged geometric hard-sphere scatters.  

 To distinguish between 
strong-darklear and  V-mediator induced interactions, we will add a $\Lambda$ or $V$ superscript respectively.
In standard BBN \cite{Wagoner:1966pv}, thermal averaging is evaluated using the ``Gamow peak" approximation, which
fails in the  $\alpha_D \ll \alpha_{EM}$ regime, so we perform the integral
in Eq.~\ref{eq:thermal-integral} directly.  

Since we require a population of light mediators to initiate di-darkleon formation via $\chi + \chi \to \,^2\!\chi + V$, 
there will also be a $V$-emission processes $^A\chi +\, ^B\chi  \to \, ^{(A+B)}\chi  + V$, in which a mediator particle is radiated off
an initial or final state particle. This process is modeled using the simple prescription 
$\langle \sigma_V v \rangle_{A,B}  = \alpha_D \langle \sigma_\Lambda v \rangle_{A,B}$ in accordance with the
 $\alpha_{EM}$ scaling of analogous visible-sector processes (e.g. $p + n \to d +\gamma$). 
We define $\Gamma^V_{A, B} \equiv n_D \langle \sigma_V v \rangle_{A, B}$ to be the rate of $V$-mediated 
darklei fusion for $A$ and $B$: $^A \chi + \, ^B\chi \to\, ^{(A+B)}\chi + \,V$, where $n_D = \rho_{DM}/m_\chi$ is the darkleon number density. 
Similarly, we define  $\Gamma^\Lambda_{A, B} \equiv n_D \langle \sigma_\Lambda v\rangle_{A,B}$ to the analogous strong-darklear process involving the same species. 
Finally, by the principle of detailed balance, the mediator-induced dissociation cross section for $V + \,^{(A+B)} \chi \to \,^A\chi + \,^B\chi$ is 
\be
\langle \sigma_V v \rangle_{(A+B)\to A,B} \equiv \frac{n_A n_B}{n_V n_{A+B}}\langle \sigma_V v \rangle_{A,B}~,~
\ee
 where  $n_i$ is the number density of the $i^{\rm th}$ species and the corresponding rate is defined $\Gamma^V_{(A+B)\to A,B} \equiv n_D \langle \sigma_V v \rangle_{(A+B)\to A,B}$.
 
%%%%%%%%%%%%%%%%%%%%%%%%%%%%%%%%%
%%%%%%%%%%%%%%%%%%%%%%%%%%%%%%%%%
%
%					Boltzmannia 
%
%%%%%%%%%%%%%%%%%%%%%%%%%%%%%%%%%
%%%%%%%%%%%%%%%%%%%%%%%%%%%%%%%%%

{\it \bf Boltzmann Equations: }
We solve the Boltzmann equations for $N$ species of darklei built up from a population of identical $\chi$ 
darkleons. In terms of number fractions $Y_i \equiv n_i/n_D$, these can be written as  
\be
\label{eq:boltz}
&&  \hspace{0.cm }\frac{dY_A}{dt}  =  \sum_{B= A+1}^{N}   \left(         \Gamma^V_{B\to A,B-A} Y_B Y_V   -   \Gamma^V_{A, B-A} Y_A Y_{B-A}      \right)
  \nonumber \\
&& \hspace{1.1cm}
  +  \sum_{B=1}^{A-1} \> \, \left(  \Gamma^V_{B,A-B}   Y_B Y_{A-B}        -  \Gamma^V_{A \to B, A-B} Y_A Y_V       \right)
  \nonumber \\
&& 
 \hspace{0.7cm} +~   \sum_{B =1}^{N} \sum_{C = 1}^{A-1} \bigl( \Gamma^{\Lambda}_{\small B+C,A-C} Y_{B+C} Y_{A-C}  - 
  \Gamma^{\Lambda}_{\small A,B}  Y_A Y_B \bigr) ~,~~~~~~~~
\ee
%\be
%\label{eq:boltz}
% \frac{dY_A}{dt}  &=&  \sum_{B= A+1}^{N}   \Gamma_{B,A} Y_B Y_V  -  \sum_{B=1}^{A-1} \Gamma_{A,B} Y_A Y_V    
%  \nonumber \\ &&
%     \hspace{-0.9cm}+ ~ n_D \sum_{B=1}^{A-1} \left(  \langle \sigma_V v \rangle_+ Y_B Y_{A-B} -    \langle \sigma_V v \rangle_- Y_A Y_V         \right)   
%  \nonumber \\
%&& 
% \hspace{-1.3cm} +~ n_D  \sum_{B =1}^{N} \sum_{C = 1}^{A-1} \bigl( \langle \sigma_\Lambda  v\rangle_+ Y_{B+C} Y_{A-C}  -\langle \sigma_\Lambda  v\rangle_-  Y_A Y_B \bigr) 
% ~,~~~
%\ee 
The first two lines of Eq.~\ref{eq:boltz} contain every $V$-mediated process that adds or removes an $A$, 
while the third line features all allowed strong-darklear processes
 $^A\chi + ^B\!\chi  \to\, ^{(A + C)}\!\chi + \,^{(B-C)}\chi $ that exchange $^{C}\chi$ darklei. 
If multiple species of darkleons are present ({\it i.e.} dark protons and neutrons), species with 
identical $A$ may have different $U(1)_D$ charges and would be tracked separately.

Figure 1 shows a density plot of the expected mass number $\langle A \rangle = \sum A^2 Y_A$ for a population of darklei with mass numbers 
$A = 1-20$ over a range of  $\alpha_D$ and $\Lambda_D$ values. We use the binding model in Eq.~\ref{eq:binding} with
parameters $a_V = 1.9\, r,  a_S = 1.3 \, r, a_P = 0.2 \, r $,$a_A = 0.6\,r$ set by
 the confinement scale through $r \equiv (\Lambda_D/\Lambda_{QCD})\gev$; the Coulomb term is $a_C = 3 \times 10^{-7} r$.
 Aside from the coulombic term, which is required to be small for a viable dark sector, we take all other inputs 
 to be of order the confinement scale; any separation of scales in this context is model dependent 
 (e.g. the range of pion interactions
 in the SM depends on quark masses).
The initial conditions assume a single-species of $\chi$ darkleons with
identical  $U(1)_D$ charges so that each process encounters the maximum  
coulomb barrier in every interaction.  A more realistic setup 
 also features a small enhancement and differentiation in some rates due to attractive interactions between oppositely charged species, but 
 we leave these details for future investigation.  Figure 2 shows the distribution of individual species for particular $\alpha_D$  and $\Lambda_D$ under
 the same assumptions and conditions used in Figure 1. 

We can roughly estimate the freeze out temperature assuming 
a fully asymmetric dark sector $n_\chi \sim (m_n \Omega_{DM}/ m_\chi \Omega_b) n_b$.
 BBD is controlled by the formation of the $^2\chi$ state via $\chi + \chi \to V +\,  ^2\chi$, 
so the relevant reaction rate scales as $\langle \sigma v \rangle \sim \alpha_D \Lambda_D^{-2} (T/m_\chi)^{1/2}$.
Evaluating $n_\chi \langle \sigma v \rangle \sim H$ yields
\be
T_{f} \sim  \gev~\left( \frac{10^{-5}}{\alpha_D}\right)^{2/3}  \left( \frac{\Lambda_D}{10^2 \, \gev}\right)^{4/3}\left( \frac{m_\chi}{10^2 \, \gev}\right)^{1/2}\!,~
\ee 
which is consistent with our results in Fig. 2.

%%%%%%%%%%%%%%%%%%%%%%%%%%%%%%%%%
%%%%%%%%%%%%%%%%%%%%%%%%%%%%%%%%%
%
%					UV Story 
%
%%%%%%%%%%%%%%%%%%%%%%%%%%%%%%%%%
%%%%%%%%%%%%%%%%%%%%%%%%%%%%%%%%%

\section{UV Completion}
\label{sec:uv}

An example UV model that generates a dark matter-asymmetry and yields BBD contains an $SU(3)_D \times U(1)_D$ dark gauge symmetry,
 $N_f$ flavors of Weyl fermions $\psi,\xi \sim \mathbf 3_{\pm 1}$, $N_f$ flavors
of  their Dirac partners $\psi^c, \xi^c \sim \overline{\mathbf 3}_{\mp 1}$, and $N_s$ flavors of scalars $\varphi \sim  \mathbf 3_0$. The lagrangian 
\be \label{eq:lag}
{\cal L} &=& 
\lambda \varphi \, \xi \psi + \lambda^\prime \varphi^{ \dagger} \, {\xi^c} {\psi^c} + m_\psi \psi\psi^c\nonumber \\ &+&  m_\xi \xi \xi^c +  M^2 \varphi^\dagger\varphi+ \mu\, \varphi \varphi \varphi  + h.c. ~~,
\ee 
 satisfies the Sakharov conditions \cite{Sakharov:1967dj} for the dark sector:  the matrices $\lambda, \lambda^\prime$ contain irreducible $CP$ violating phases, 
 the trilinear scalar interaction explicitly violates
a global DM ``baryon" number under which $\varphi\sim -2$ and  $ \xi, \psi  ( \psi^c, \xi^c)   \sim \pm1 $, and the scalars 
$\varphi, \varphi^\dagger$ can decay out of equilibrium in the early universe. 
All gauge and flavor indices in the couplings and masses of Eq.~\ref{eq:lag} have been suppressed.
Interference between tree and loop decay-diagrams induce $C$ and $CP$ violation and yield a matter asymmetry in the dark 
sector following standard methods \cite{Kolb:1990vq}. 
  
 A confining phase transition occurs at  $T \sim \Lambda_D$,
 below which the stable confined  darkleons  come in different species  
$ \chi_{_{3}} =  (\psi \psi \psi),~ \chi_{_{1}}=  (\psi \psi \xi), ~\chi_{_{-1}} = (\psi \xi \xi)$ and $ \chi_{_{-3}}= (\xi \xi \xi)$
which carry $U(1)_D$ charges $3,1, -1,$ and$ -3$, respectively.
Each combination of states can fuse to form 
 darklei; however, the initial condition is no longer an identical population of $\chi$, but a  distribution of distinct $\chi_i$ with
 both attractive and repulsive interactions during BBD. 
 
 For $m_{\psi,\xi} \ll \Lambda_D$,the confinement breaks an approximate $SU(N_f) \times SU(N_f)$ chiral symmetry under which 
 $\chi (\xi)$ and $\psi^c (\xi^c)$ can be independently rotated. In this regime, the IR spectrum contains pseudo-goldstone bosons (dark-pions) with mass-squared 
 proportional to  $m_{\psi, \xi}$ in analogy with low-energy QCD. The $U(1)_D$ neutral
  pions (e.g. $\psi \psi^c$ states) decay to the visible sector via kinetic mixing, while charged pions
  (e.g. $\psi \xi^c)$ are matter-symmetric and annihilate to visible states.

\section{Discussion}
\label{sec:conclusion}

In this paper we have conservatively shown that, in broad regions of viable parameter space,
 asymmetric dark matter models with
 nonabelian confinement efficiently produce {\it darklei}, dark-sector  nuclei, in the early universe.
  Unlike visible BBN, which involves several (possibly anthropic) coincidences that impede the synthesis of heavier nuclei, {\it darkleosynthesis}
 can be highly efficient and proceed to high mass states. Indeed, the light mediator particle that enables dark-deuterium
 formation can have a smaller coupling than $\alpha_{EM}$, so coulomb-like barriers 
 that prevent the formation of high-$Z$ elements are {\it exponentially} less inhibiting.
  Furthermore, the dark confinement-scale, which sets the binding energy scale, can be much larger than typical binding energies in 
 the visible sector, so larger species are more tightly bound.
Finally, nucleosynthesis in the dark sector 
can begin at higher temperatures and freeze-out later than visible-BBN, thereby extending the duration 
of reactions.
 
For simplicity we have only computed the yields for each species and ignored
other novel features of the darklear isotope distribution at late times, which we leave for future work. 
Our approach does not attempt to understand the precise details
of dark-nuclear interactions; we model formation rates with geometric cross-sections and binding energies
with the liquid drop model merely to demonstrate the hitherto overlooked possibility of dark-nucleosynthesis; a more realistic
understanding of confined dynamics would allow for a more detailed investigation.  

Since darkleons can be significantly lighter than large-$A$ darklei,
 their particle-antiparticle annihilations (or those of  their constituents) can still be efficient 
 at high temperatures, while assembling large-$A$ composites at later times. 
Thus, naive overclosure need not be  a limitation for heavy, thermally produced DM. 
 Furthermore, as with many nuclei in the visible sector, BBD can yield composites darklear states with spin $>$
 $3/2$, so scattering at direct detection experiments may offer novel directional signatures with multiple species and form factors that respond 
 differently to various target materials. Since no other known mechanism can yield interacting, higher spin particles, discovering 
 dark matter with spin $> 3/2$ would be smoking gun evidence of darkleosynthesis.
 
 We note in passing that  darkleosynthesis may be an extended process that continues until late times, 
perhaps even in dark matter halos in the present epoch to realize a novel form of self interacting dark matter whose
indirect detection signatures arise from visible emission during dark fusion. 
Furthermore, if metastable excited states are long lived, exothermic darklear scattering at late times may facilitate
 ``dark disk" formation \cite{Fan:2013yva,Fan:2013tia} in dark matter halos subject to the cosmological limits on
  dark matter interactions with relativistic species \cite{Cyr-Racine:2013fsa,Buckley:2014hja}.

{\it Acknowledgments}
We thank Maxim Pospelov,
 Philip Schuster,  Flip Tanedo, Brian Shuve, Natalia Toro, and Neal Weiner for helpful conversations. 
The research of KS is supported
in part by a National Science and Engineering Research
Council (NSERC) of Canada Discovery Grant. This research was supported in part by Perimeter Institute for Theoretical Physics. The Perimeter Institute for Theoretical Physics is 
 is supported by the Government of Canada through Industry Canada
and by the Province of Ontario.
 
\bibliographystyle{apsrevM}
\bibliography{BBD-Draft}

\end{document}